# Exactly Solvable Schrödinger equations with Singularities: A Systematic Approach to Solving Complexified Potentials (part1)

Jamal Benbourenane
D.Benbourenane@Chatham.edu
Chatham University, Pittsburgh, PA 15232
January 9, 2023

This paper gives a new perspective on how to solve the second-order linear differential equation written in normal form. Extending the argument of the potential to a complex number leads to solving exactly the Schrödinger equation when the potential is complex using the factorization method. This method leads to solving two Riccati nonlinear equations and by constructing the only possible superpotential, the factorization method gives the eigenvalues and eigenfunctions in closed form for potentials satisfying the shape invariance property.

Extending the potential to the complex argument has led to discovering new exactly solvable ones. In this first part, the basic superpotentials are divided into different groups, each group contains the superpotentials that share common terms. All of the already known solvable real potentials will fall into this category and are derived as special cases. This set of exactly solvable complexified potentials has already uncovered some of the properties of quantum mechanics, like the tunneling effect through the forbidden region, happening with high probabilities between multiwells, bound states in the continuum (BIC), and other properties. These results have potential applications in all fields of sciences, from physics, chemistry, biology, etc., where the eigenvalue problem plays an important role.

## 1 Introduction

The goal of this paper is to find the general solution to the second-order linear differential equation

$$-\frac{d^2y}{dx^2} + V(x)y = 0 \qquad (1)$$

where $x$, the independent variable, is real in the whole paper, and where the coefficient $V(x)$ will be constructed and is allowed to be complex-valued, as well as having singularities or being periodic, aperiodic, symmetric, or asymmetric, with no restriction. We already know that if one solution $y_1$ exists, then we can find the second solution $y_2$ by reduction of order, $y_2 = y_1 \int \frac{1}{y^2}$, and the general solution is then written as a linear combination of the two in the form $y = c_1 y_1 + c_2 y_2$. However, in those simplest cases of a constant $V(x)$, the solutions are found to be trigonometric or hyperbolic (depending on the constant itself), obtained not by a formula, but by a mere guess. So, if a simple problem like this one requires guessing and constructing a solution, what can we expect in the case of more difficult problems, can we still guess the solutions? When extending the problem to a more general equation where our equation (1) becomes a special case, it leads us to solve a more general case in the form of a second-order linear partial differential equation, called the time-dependent Schrödinger equation

$$i\frac{\partial y}{\partial t} = -\frac{\partial^2 y}{\partial x^2} + V(x)y \qquad (2)$$

where $y = y(x,t)$ is a function in two variables and $V$ is a complex coefficient called a potential function that depends only on the real variable $x$. So, at first glance, it looks like we are trying to solve a rather more difficult problem than the original one by assuming a complex potential and by adding another term on the left side, however, we will see that on the contrary, this approach will help in providing us with two Riccati equations that we can use to construct the only possible potentials that solve (2). This equation reduces to the second-order linear differential equation (1) that we started with if $y$ depends only on $x$ only. So, not only we will be able to solve the second-order linear partial differential equation (1), but also be able to solve the second-order linear differential equation (2), and the nonlinear Ricatti differential equation.

By taking the solution to be a product of two separable functions in the form

$$y(x,t) = h(t)\Psi(x), \qquad (3)$$

we obtain

$$i\frac{h'(t)}{h(t)} = -\frac{\Psi''(x)}{\Psi(x)} + V(x) \qquad (4)$$

The two quantities of the equation (4) are independent, the left side depends on $t$, while the right side depends on $x$, therefore, they should be equal to a common constant $\lambda$, called an eigenvalue. This leads to two equations

$$-\Psi'' + V(x)\Psi = \lambda\Psi \qquad (5)$$

and

$$i\frac{h'(t)}{h(t)} = \lambda \qquad (6)$$

and the solution to (2) comes from (3) and (5) and is given by

$$y(x,t) = e^{-i\lambda t}\Psi(x) \qquad (7)$$

The imaginary number on the left side of the equation 2 is just a real constant, historically chosen to be the measurement of the energy, so that equation (6) will have a time-dependent phase factor where the magnitude of $y(x,t)$ is independent of the real-time $t$, i.e $|y(x,t)|^2 = |\Psi(x)|^2$.

This equation (5) is the celebrated time-independent Schrödinger equation in one-dimensional space from quantum physics. Therefore, the terminology from physics, such as potential, wavefunctions, eigenfunctions, eigenenergies, and other terms, will be used interchangeably throughout the paper.

The solution $y$ to the second-order linear differential equation (1) will be found when the solution $\Psi$ corresponding to the zero-eigenvalue $\lambda$ in equation (5) is found. However, the question that puzzled mathematicians and physicists is how to find the formulae in closed forms to the solutions $\Psi$ and the eigenvalue $\lambda$ of the Schrödinger equation (5), and which potentials $V(x)$ allow such solutions. The problem of asking to find three unknowns from a single equation seems to be a difficult task. It is well known that only a handful of such potentials with exact solutions are found to this day.

Four centuries ago, mathematicians faced the same difficult question about the solutions of another second-order equation, but at that time it was a second-order polynomial equation $x^2 + a^2 = 0$. Putting the original equation $-y'' + V\, y = 0$ against this one, they look almost similar, isn't it? At that time, they could not grasp the idea that such solutions could exist. Later, when its existence was accepted, they attached to it a derogatory name of an imaginary number. The term was coined by René Descartes who was skeptical about its usefulness.

I am going to show in this paper and the other papers in the pipeline that these imaginary numbers are very useful. They can help us break the deadlock.

After Schrödinger introduced his equation (5) in quantum mechanics, the harmonic potential and Coulomb potential were solved. Other authors, like Morse, Rosen, Eckart, Scarf, Pöschl, and Teller could come up with the exact solutions to a few other potentials. The last exactly solvable potential in analytic form was published in 1958.

The factorization method is the natural way to solve the second-order differential equation by writing the left side of the Schrödinger equation into two non-commutative products of linear first-order differential operators. It is among the most successful method for evaluating the exact eigenvalues and eigenfunctions. Even though the factorization method was known to Schrödinger [31] and others, it was until 1951 that Infeld and Hull [20] came up with a concise way of describing the method. Then, in 1981, Witten introduced supersymmetry (SUSY) in non-relativistic quantum mechanics [34]., and was followed by Gendenshtein in 1983 [18], who introduced the concept of shape invariance property that some potentials possess. It was later shown that all known solvable potentials are shape-invariant. This same idea was used by the author from 2020 papers onward [7][6][2][3][4][5].

In this paper, I will introduce the new solvable potentials $V(x, a_0)$ by their superpotential $W(x, a_0)$, in form $V(x, a_0) = W^2(x, a_0) - \frac{d}{dx} W(x, a_0)$, where the parameter $a_0$ comes from the sequence $\{a_k\}$ satisfying the shape-invariant condition, and $W$ is called the superpotential. For simplicity and to avoid crowdedness, I will omit writing in full the potential $V$, which is lengthy for many potentials, and will only provide its companion, the superpotential $W$.

Since there is a long list of solvable potentials, I will start by naming these superpotentials with ascending numbers when plotting them. Also, I will group them into different groups according to the shared properties inherited from their superpotentials. Each group can contain one or more solvable potentials. In this paper I have selected only 13 groups, the other groups of new superpotentials will be discussed in different articles.

When we speak about exact solutions to differential equations, we mean that the solution will be given explicitly in a closed form for both the eigenvalues and the eigenfunctions. In the case of the Schrödinger equation, the formula obtained for the eigenvalues is algebraic in its form and depends on the sequence of numbers $\{a_k\}$, which is challenging to construct. While the formula of the eigenfunctions is a recursive one, which involves the employment of simple operations, like addition, subtraction, multiplication, division, differentiation, and integration. Any solution given implicitly is not allowed. We also need to make a distinction between exactly solvable and analytically solvable, they are different concepts. In this paper, there will be no approximation, truncation, or asymptotically equivalent expression but only exact formulae.

The eigenvalues we seek to find, in either real or complex potentials, are ordered positive real numbers. It has been a long debate about the possibility of having real spectra when the potential is complex. The answer to this question will be given with concrete examples of constructed potentials (real or complexified) having real eigenvalues which will finally settle the debate.

We know from the literature that the exactly solvable potentials were shown to be shape-invariant and those formulae were obtained with the independent variable $x$ being extended by scaling the variable $x$ using the transformation $x \to px$.

In this paper, all the parameters appearing in the superpotential $W(x, a_0)$ are defined to be real. It is assumed that we start from a real potential, then we complexify the argument, the coefficient(s), or both. I will start by extending the solvable potentials by complexifying the independent variable $x$ in $W$ with the change of variable $x \to px + iq$, such potentials are said to be obtained by complexification. When we multiply one or more of the real coefficients found in $a_k$ by the imaginary number $i$, such potential will be called complexified potential. It is important to make this distinction since both are complex potentials, however, defined differently. If both types of transformations are applied to a real potential we say that we have a complexification of the complexified potential. For example $W(x, a_0) = -A_0 \cot x + B_0$ is the real superpotential of the Rosen-Morse potential, if I multiply the coefficient $B_0$ by $i$, the new superpotential is $W(x, a_0) = -A_0 \cot x + iB_0$ and the new potential will be called the complexified potential. If we complexify $x$ then $W(x, a_0) = -A_0 \cot(px + iq) + iB_0$, and the corresponding potential is said to be the complexification of the complexified potential.

The superpotentials are taken in a way that the real part of the partner potentials $V$ always has at least one minimum below the real axis, so in all given plots, it is assumed that the real part of the complex potential has a minimum. The idea behind the complexification of the potential $V$ is, firstly, avoiding the singularities inherited by a potential through the superpotential $W$, by going around each singular point in the direction parallel to the imaginary axis; and, secondly, enabling to deal with periodic superpotentials and the possibility of obtaining multiwells potentials. But, in general, this idea works for all types of potentials even if there is no singularity. It is just an extension from the real space to the complex plane of the argument of the potential by a simple linear transformation.

In the next section, I will give a brief introduction to the factorization method, also called the supersymmetry technique. Then, in section 3, I will cover the shape invariance property with the algebraic formula of the eigenvalues and the recursive formula of the eigenfunctions, which will be the focal points for all proposed potentials. In section 4, I introduce the groups of solvable complexified potentials through their superpotentials by defining the corresponding sequence $\{a_k\}$ satisfying the shape invariance property and then deriving the formula for the eigenvalues. The eigenfunctions are given recursively and depend closely on the sequence $\{a_k\}$, their plots will be given alongside the potential and eigenvalues. Having these eigenfunctions written in a closed form, now I turn to the possible applications in quantum mechanics, after normalizing them, and making sure they vanish at the two boundaries of the real domain. Hence, some of the hidden properties never seen before are uncovered, like bounded states in the continuum (BIC), resonant states, states defined in multiwells potential, and states in periodic potentials. These potentials will include, as special cases, some of the well-known solvable potentials. In section 5, I conclude by summarising the discoveries and their implications.

For those familiar with the factorization method they can skip the next two sections and go directly to the main results in section 4.

## 2  Factorization Method and Supersymmetry

Given a potential $V_-(x)$, we seek to build a partner potential $V_+(x)$, where these two potentials have the same energy eigenvalues, except for the ground state.

These partner potentials are defined by
$$V_\pm(x) = W^2(x) \pm W'(x) \tag{8}$$

where $W(x)$ is called the superpotential.

The associated Hamiltonians to these partner potentials are given by
$$H_- = A^\dagger A, \quad H_+ = AA^\dagger \tag{9}$$

where
$$A^\dagger = -\frac{d}{dx} + W(x), \quad A = \frac{d}{dx} + W(x) \tag{10}$$

The time-independent Schrödinger equation in 1-D with eigenstate $E$ and potential $V(x)$ is defined by
$$\left(-\frac{d^2}{dx^2} + V(x)\right)\Psi = E\Psi \tag{11}$$

and the two Hamiltonians associated with the Schrödinger equation are written in the form
$$H_- = -\frac{d^2}{dx^2} + V_-(x), \quad H_+ = -\frac{d^2}{dx^2} + V_+(x), \tag{12}$$

The two Hamiltonians are both positive semi-definite operators, so their energies are greater than or equal to zero.

For the Hamiltonian $H_-$, we have
$$H_-\Psi_0^{(-)} = A^\dagger A \Psi_0^{(-)} = E_0^{(-)}\Psi_0^{(-)}$$

so that
$$H_+(A\Psi_0^{(-)}) = E_0^{(-)}(A\Psi_0^{(-)}).$$

Similarly, for the Hamiltonian $H_+$
$$H_+\Psi_0^{(+)} = AA^\dagger \Psi_0^{(+)} = E_0^{(+)}\Psi_0^{(+)}$$

and multiplying the left side by $A^\dagger$ we obtain

$$H_-(A^\dagger \Psi_0^{(+)}) = E_0^{(+)}(A^\dagger \Psi_0^{(+)}). \tag{13}$$

The eigenfunctions of the two Hamiltonians and their exact relationships will depend on whether the quantity $A\Psi_0^{(-)}$ is zero or nonzero, i.e. if $E_0^{(-)}$ is zero or nonzero, which means an unbroken supersymmetric system or a broken one. For a full discussion of these cases, see the references[32] [33].

Thus, here I will consider only the case of unbroken supersymmetry, where $A\Psi_0^{(-)} = 0$. In this case, this state has no SUSY partner since the ground state wavefunction of $H_-$ is annihilated by the operator $A$, in which case $E_0^{(-)} = 0$.

It is then clear that the eigenstates and eigenfunctions of the two Hamiltonians $H_-$ and $H_+$ are related by (for $n = 0,1,2,\dots$)

$$E_0^{(-)} = 0, \quad E_n^{(+)} = E_{n+1}^{(-)}, \tag{14}$$

$$\Psi_n^{(+)} = \left(E_{n+1}^{(-)}\right)^{-1/2} A \Psi_{n+1}^{(-)} \tag{15}$$

$$\Psi_{n+1}^{(-)} = \left(E_n^{(+)}\right)^{-1/2} A^\dagger \Psi_n^{(+)}. \tag{16}$$

This process is obtained, as in the case of the harmonic oscillator, by applying the creation and annihilation operators.

By knowing $W(x)$, then the ground state wavefunction $\Psi_0^{(-)}$ can be expressed by

$$\Psi_0^{(-)} = N \ e^{-\int W(x)dx} \tag{17}$$

where $N$ is the normalized constant,

Importantly, note that $\Psi_0 = \Psi_0^{(-)}$ is a solution to the second-order differential equation (1), therefore, solving the second-order differential will only depend on the constructed $W(x)$.

In quantum mechanics, the bound state wavefunctions must converge to zero at the two ends of its real domain interval, therefore the two statements $A\Psi_0^{(-)} = 0$ and $A^\dagger \Psi_0^{(+)} = 0$ cannot be satisfied at the same time, and so only one of the two ground state energies, $E_0^{(-)}$ and $E_0^{(+)}$, can be zero, while the other bound state energy will be positive. So, we will define by convention $W$ such that $\Psi_0$ is normalized with

$$E_0^{(-)} = 0, \ E_1^{(-)} = E_0^{(+)} > 0. \tag{18}$$

## 3   Shape Invariance Property

As we have seen in the last section, to solve the Schrödinger equation (11) having a potential $V(x)$ using the method of supersymmetry, we are required to construct the superpotential $W$. This is equivalent to solving the Riccati differential equation (8). However, the Riccati equation is also an unsolved equation except for a limited number of them. But, the Riccati equation happens to be more tractable than the original Schrödinger equation for some potentials with a particular geometric property of the shape of the partner potential as can be seen in the known solvable potentials.

We will define a potential to have a shape-invariant property if the dependence on $x$ of the partner potentials is similar and only differ on some parameters appearing in their expressions. This similarity is described by the following relation :

$$V_-(x, a_1) + h(a_1) = V_+(x, a_0) + h(a_0) \qquad (19)$$

where the parameter $a_1$ depends on $a_0$, i.e. $a_1 = f(a_0)$, and then, $a_2 = f(a_1) = f^2(a_0)$, and by recurrence $a_k = f^k(a_0)$ where $a_0 = (A_0, B_0, ..) \in \mathbb{R}^m$, and $f: \mathbb{R}^m \to \mathbb{R}^m$, is called a parameter change function, see [1][9][12][14][17].

In the shape invariance method, the parameters of the superpotentials are changed, but the partner potentials have similar shapes and differ by a constant.

It is no more a challenge to find solutions to the equation (19) of the shape-invariant condition, as I have demonstrated in [7] and others. Indeed, It was noted in [9] that In the past, researchers had found a list of additive shape-invariant potentials, mostly by trial and error, however, in this paper and future papers, the superpotentials are found rigorously, and methodically, which helped in producing, so far, a long list of solvable potentials.

From now on, I will consider the potential $V(x)$ to be shape-invariant potential. The ground state is given by,

$$\Psi_0(x, a_0) \propto e^{-\int W(x,a_0)} \qquad (20)$$

The two supersymmetric partner potentials $V_-(x, a_1)$ and $V_+(x, a_0)$ have the same dependence on $x$, up to the change in their parameters, and their Hamiltonians $H_-(x, a_1)$ and $H_+(x, a_0)$ will only differ by a vertical shift given by $C(a_0) = h(a_1) - h(a_0)$, from the equation

$$V_+(x, a_0) + h(a_0) = V_-(x, a_1) + h(a_1) \qquad (21)$$

where the partner potentials are defined by

$$V_-(x, a_1) = W^2(x, a_1) - W'(x, a_1), \qquad (22)$$
$$V_+(x, a_0) = W^2(x, a_0) + W'(x, a_0) \qquad (23)$$

The shared ground state wavefunction for these two Hamiltonians is given by

$$\Psi_0^{(+)}(x, a_0) = \Psi_0^{(-)}(x, a_1) \propto e^{-\int W(x,a_1)} \qquad (24)$$

The first excited state $\Psi_1^{(-)}$ of $H_-(x, a_1)$, omitting the normalization constant, is given by

$$\Psi_1^{(-)}(x, a_0) = A^\dagger(x, a_0)\Psi_0^{(+)}(x, a_0) = A^\dagger(x, a_0)\Psi_0^{(-)}(x, a_1). \qquad (25)$$

The eigenvalue associated with this Hamiltonian is
$$E_1^{(-)} = C(a_0) = h(a_1) - h(a_0) \tag{26}$$

The eigenvalues of the two Hamiltonians $H_+$ and $H_-$ have the same eigenvalues except for the additional zero energy eigenvalue of the lower ladder Hamiltonian $H_-$. They are related by

$$E_0^{(-)} = 0, \quad E_{n+1}^{(-)} = E_n^{(+)}, \tag{27}$$

$$\Psi_n^{(+)} \propto A\Psi_{n+1}^{(-)}, A^\dagger \Psi_n^{(+)} \propto \Psi_{n+1}^{(-)}, \quad n = 0,1,2,\ldots \tag{28}$$

where this procedure has been iterated to construct a hierarchy of Hamiltonians

$$H_\pm^{(n)} = -\frac{d^2}{dx^2} + V_\pm(x, a_n) + \sum_{k=0}^{n-1} C(a_k) \tag{29}$$

and then derived the $n^{th}$ excited eigenfunction and eigenvalues by

$$\Psi_n^{(-)}(x, a_0) \propto A^\dagger(x, a_0) A^\dagger(x, a_1) \ldots A^\dagger(x, a_n) \Psi_0^{(-)}(x, a_n) \tag{30}$$

$$\begin{aligned} E_0^{(-)} &= 0, \\ E_n^{(-)} &= \sum_{k=0}^{n-1} C(a_k) = \sum_{k=0}^{n-1} h(a_{k+1}) - h(a_k) \\ &= h(a_n) - h(a_0), \text{ for } n \geq 1. \end{aligned} \tag{31}$$

where $a_k = f(f(\ldots f(a_0))) = f^k(a_0)$, $k = 0,1,2,\ldots, n-1$.

Therefore, knowing the superpotential not only we know the potential, but also its ground state and from the algorithm above, the whole spectrum of the Hamiltonian $H_-$ ($H_+$ as well) can be derived by the supersymmetry quantum mechanics method.

From my prior knowledge of potentials having exact solutions which I have been investigating in the last three years by using the factorization method, I have divided them into groups of superpotentials sharing the same type of forms. The first of these groups will be discussed in this section. There is no better way to showcase these new potentials and make them easily accessible to everyone, than by summarising the results in plots of the real part of selected potentials graphed together with the square of the magnitude of the eigenfunctions, and their corresponding nonnegative eigenvalues.

Using the preferred CAS software, these are the steps to follow in the order given using the exact formulas:

- Start by defining $W(x, a_0)$ given as an ansatz

- Compute $V(x, a_0) = W^2(x, a_0) - W'(x, a_0)$

- Extract the real part of $V(x, a_0)$

- Write the formula for the sequence $\{a_k\}$

- Write the formula of the eigenvalues $E_n$ obtained recursively from $a_n$

- Compute $\Psi_0^{(-)}(x, a_k)$ obtained from (24)

- Compute $\Psi_n^{(-)}(x, a_0) \propto A^\dagger(x, a_0) A^\dagger(x, a_1) \ldots A^\dagger(x, a_n) \Psi_0^{(-)}(x, a_n)$, recursively.

The exact computation of the eigenfunctions cannot be done by hand for higher excited states, especially more so in the future groups of solvable potentials. There are few cases where one of the eigenfunctions, let's say, $\Psi_1$, is not normalizable, then we use instead the complement function $\Psi_2 = \Psi_1 \int \frac{1}{\Psi_1^2}$ if it is normalizable. If it is not normalizable, then take a linear combination of the two and make sure that the new eigenfunction is smooth by stitching the two functions together at the right points. For symmetric potentials the right point will be on the y-axis, which is the case of most complexified potentials. The figures in section 4 are the plots that include the following: the real part of $V(x, a_0)$ just call it $V(x)$, for short, the square of the moduli of the eigenfunctions, $|\Psi_k|^2$, normalized, and the eigenvalues $E_k$, $k = 0, \ldots, n$, where $n$ represents the number of some of the first allowed bound states. Usually, seven excited are enough to describe a system from what I have experienced so far.

## 4   Exactly solvable potentials given by their superpotentials

After introducing the factorization method for the shape invariance property, I am ready to list the first part of the solvable potentials by providing the superpotentials $W$, the formula for the sequence $\{a_k\}$, the formula for the energy $E_n$ and the formula for the wavefunctions. In this first part, I will try to include all the solvable potentials obtained by the shape invariant method with their more general extensions and also include some newly found solvable potentials. These potentials are divided into groups that share the same properties found in their superpotentials. All parameters $p, q, r, s$ and $A_0, B_0$ are real, but for the sake of computational implementation in CAS, It is assumed that these coefficients are positive. To investigate the cases when one or more of these parameters are negative, we need to replace the corresponding parameter with their opposite signs and obtain a new superpotential where all parameters are positive. It could be done as a follow upp paper.

The coming parts to this sequel of solvable potentials will be completely new and have never been proposed before in the literatuure with very interesting results.

As I mentioned before, to find the potential with a real argument we need to set $q = 0$, or let it approach zero, $q \to 0$, in a singular superpotential, if we want to study the behavior of the solutions near the singularity. This way we can calibrate the numerical methods around these singularities to avoid catastrophic results

The paper is more concerned with the more general case of the complexification of a potential that possesses real eigenvalues for physics implementation. So, let's start with the first group of superpotentials, listing 13 of them with a total of 24 superpotentials.

### 4.1   Group 1: Superpotentials with linear in $x$ (shifted harmonic oscillator complexified).

### 4.1.1 Superpotential $W(x, A_0, B_0) = A_0(px + iq) + B_0$

This is the well know harmonic oscillator but this time the argument is complexified.

$$V_0(x, A_0, B_0) = B_0^2 - A_0 p - A_0^2 q^2 + 2A_0 B_0 px + A_0^2 p^2 x^2 + 2i(A_0 B_0 + A_0^2 px)q$$

We consider the real part of this potential for its physical meaning.
$$V(x) = B_0^2 - A_0 p - A_0^2 q^2 + 2A_0 B_0 px + A_0^2 p^2 x^2$$
$$= (B_0 + pA_0 x)^2 - A_0 p - A_0^2 q^2$$

the sequence $\{a_k\}$, in this part 1 paper, will only depend on two parameters, $a_k = (A_k, B_k)$.

For the shifted harmonic oscillator
$$A_k = A_0, \quad B_k = B_0$$
and the first eigenenergy is defined by
$$E_1^{(-)} = E_0^{(+)} = C(a_0) = 2pA_0 - (pA_0 + B_0^2) + (pA_1 + B_1^2) \tag{32}$$
$$= 2pA_0 \tag{33}$$

By recurrence, we have
$$C_k = 2pA_k + c_k - c_{k+1}, \quad k = 0,1,2,\ldots \tag{34}$$
where,
$$c_k = -(pA_k + B_k^2) = -(pA_0 + B_0^2) \tag{35}$$

Thus, for all $n = 1,2,\ldots$, the $n^{th}$ eigenenergy or eigenvalue is given by
$$E_n = E_n^{(-)} = \sum_{k=0}^{n-1} C(a_k) = 2pnA_0 + c_0 - c_n \tag{36}$$

So, for any $n \geq 0$,
$$E_n = 2pA_0 n$$

The ground wavefunction is obtained using (24)
$$\Psi_0^{(-)}(x, a_1) \propto e^{-\int W(x, a_1)} \tag{37}$$

The first excited state wavefunction is
$$\Psi_1(x, a_0) = \left(-\frac{d}{dx} + W(x, a_0)\right) \Psi_0(x, a_1) \tag{38}$$

and the other eigenfunctions are obtained by recurrence using the formula

$$\Psi_n(x, a_k) = (-\frac{d}{dx} + W(x, a_k))\Psi_{n-1}(x, a_{k+1}), \qquad \text{for} k = 0,..,n-1. \tag{39}$$

The number of bound states $n$ for the nonnegative potential $V$ following the physical constraints on the energies $E_n$ (53): for all $n \geq 1$
$$0 \leq E_{n-1} < E_n \tag{40}$$

Here Fig. 1 describing the complexified asymmetric harmonic oscillator.

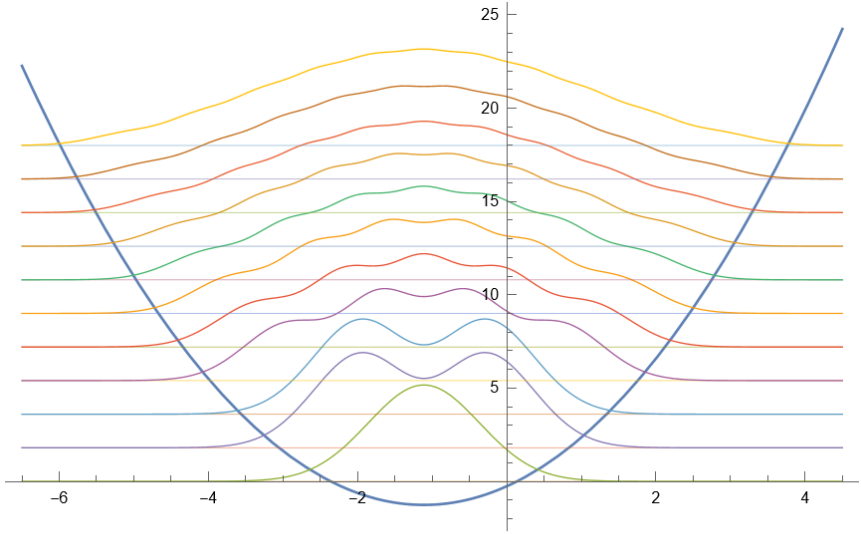

Fig 1. Potential G101, Complexified harmonic oscillator, A0=3;B0=1;p=0.3;q=0.2

### 4.1.2 Superpotential $W(x, A_0, B_0) = A_0(px + iq) + iB_0$

This is the well know harmonic oscillator but this time the argument is complexified as well as it's constant.

$$V_0(x, A_0, B_0) = -B_0^2 - A_0 p - A_0^2 q^2 - 2A_0 B_0 + 2A_0 B_0 px + A_0^2 p^2 x^2 + 2i(A_0 B_0 + A_0^2)pqx$$

The real part of this potential is
$$V(x) = -B_0^2 - A_0 p - A_0^2 q^2 - 2A_0 B_0 + 2A_0 B_0 px + A_0^2 p^2 x^2$$
$$= (B_0 + pA_0 x)^2 - 2B_0^2 - A_0 p - A_0^2 q^2$$

the sequence $\{a_k\}$ is defined by
$$A_k = A_0, \quad B_k = B_0$$

and the energies by
$$E_1^{(-)} = E_0^{(+)} = C(a_0) = 2pA_0 - (pA_0 - B_0^2) + (pA_1 - B_1^2) \tag{41}$$
$$= 2pA_0 \tag{42}$$

By recurrence, we have
$$C_k = 2pA_k + c_k - c_{k+1}, \quad k = 0,1,2,\ldots \tag{43}$$

where,
$$c_k = -(pA_k - B_k^2) = -(pA_0 - B_0^2) \tag{44}$$

Thus, for all $n = 1,2,\ldots$, the $n^{th}$ bound energy level is given by
$$E_n = E_n^{(-)} = \sum_{k=0}^{n-1} C(a_k) = 2pnA_0 + c_0 - c_n \tag{45}$$

So, for any $n \geq 0$,
$$E_n = 2pA_0 n$$

Fig. 2 is the symmetric complexified harmonic potential.

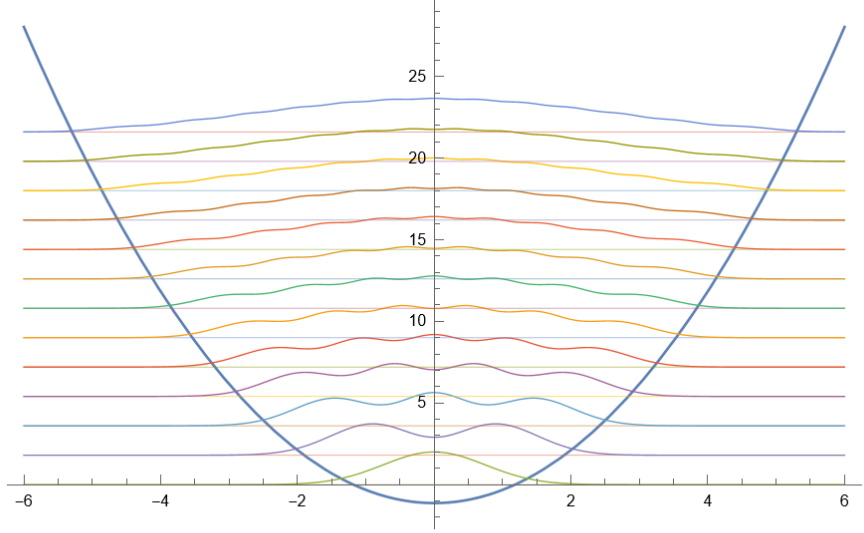

Fig 2. Potential G102. Complexified shifted harmonic oscillator with complexed constant,

A0=3;B0=0.1;p=0.3;q=0.2

## 4.2 Group 2: Superpotentials with a reciprocal of $x$ and a constant (Coulomb complexified potential).

### 4.2.1 Superpotential $W(x, A_0, B_0) = A_0 - \frac{B_0}{px+iq}$

There is an interesting consequence when we complexify the coulomb potential. It can be seen that this complex potential has two double wells not found in the real potential, see Fig. 3.

The sequence $a_k = (A_k, B_k)$ is given by
$$A_k = \frac{A_o B_0}{B_0 + kp}, B_k = B_0 + kp$$

The energy values are given by
$$E_n = A_0^2 - A_n^2$$

The real part of the complex potential is
$$V(x) = A_0^2 + \frac{B_0(B_0-p)(-q^2+p^2x^2)}{(q^2+p^2x^2)^2} - \frac{2A_0 B_0 px}{q^2+p^2x^2}$$

There are infinitely many bound energy levels converging to $A_0^2$. For the chosen parameters in Fig. 3, it is observed two minima (the other minima is too deep, not shown here)

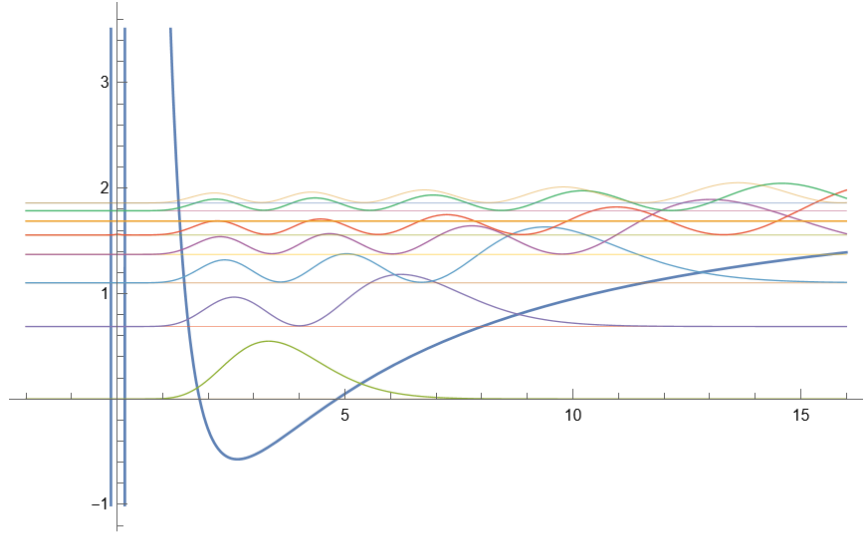

Fig 3. Potential G201, Complexified Coulomb potential with nth bound state energies converging and not exceeding A0$^2$, A0=1.5;B0=10;p=2;q=0.3

### 4.2.2 Superpotential $W(x, A_0, B_0) = iA_0 + \frac{B_0}{px+iq}$

When complexifying the constant of the coulomb's superpotential.
The sequence $a_k = (A_k, B_k)$ is given by
$$A_k = \frac{A_o B_0}{B_0 - kp}, B_k = B_0 - kp$$

The energy values are given by
$$E_n = -A_0^2 + A_n^2$$

The real part of the complex potential is
$$V(x) = -A_0^2 - \frac{B_0(B_0+p)q^2}{(q^2+p^2x^2)^2} + \frac{2B_0(B_0+p+2A_0q)}{q^2+p^2x^2}$$

There is a finite number of bound states and we can observe bound states in the continuum, BICs, as well as tunneling, see Fig. 4. The number of bound states cannot exceed $\frac{2B_0-p}{2p}$.

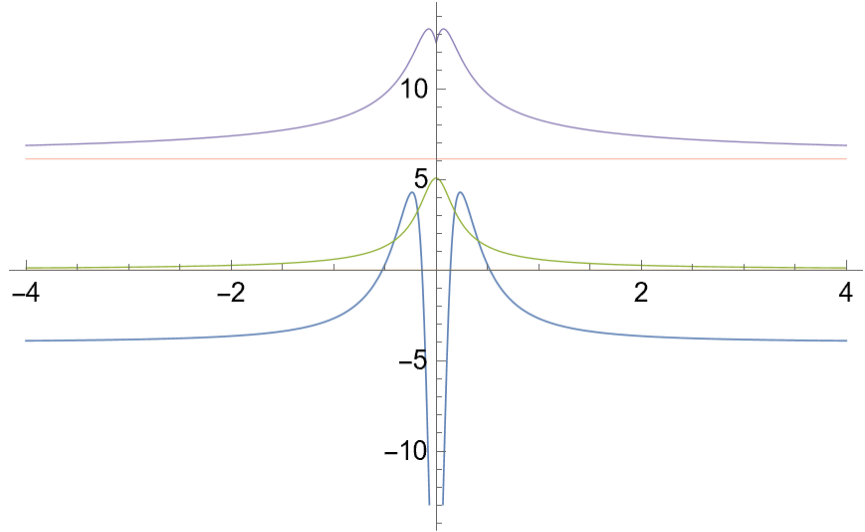

Fig 4. Potential G202. Complexification of the Coulomb potential.

A two-level system with one BIC state and tunneling of the ground state;

A0=2;B0=0.35;p-0.57;q=0.1

### 4.3  Group 3: Superpotentials with $x$ and a reciprocal of $x$ (3-D Oscillator complexified potential).

#### 4.3.1  Superpotential $W(x, A_0, B_0) = A_0(x + iq) - \frac{B_0}{px+iq}$

There are three cases.
Case 1: The sequence $a_k = (A_k, B_k)$ is given by
$$A_k = (-1)^k A_0, B_k = B_0 + kp$$

The eigenvalues are given by
$$E_n = 2A_0 p n - 2A_0 B_0 + 2A_n B_n + (-A_0 + A_n)p$$

The only case where the eigenvalue is greater or equal to zero when all parameters are positive is the ground state. So, this is a one-state system. But, for mathematical reasons I have given the formula here for all eigenvalues.

Case 2: The sequence $a_k = (A_k, B_k)$ is given by
$$A_k = A_0, B_k = (-1)^k B_0$$

The energy values are given by
$$E_n = 2A_0 p n - 2A_0 B_0 + 2A_n B_n + (-A_0 + A_n)p$$

The real part of the complex potential is

$$V(x) = 2A_0B_0 - A_0p + \frac{B_0^2}{p^2r^2} + \frac{B_0}{pr^2} + A_0^2p^2r^2$$

Case 2: The sequence $a_k = (A_k, B_k)$ is given by
$$A_k = A_0, B_k = (-1)^k B_0$$

The energy values are given by
$$E_n = 2A_0pn + 2A_0B_0 - 2A_nB_n + (-A_0 + A_n)p$$

The real part of the complex potential is
$$V(x) = -2A_0B_0 - A_0p - A_0^2q^2 - \frac{2B_0q^2(B_0-p)}{(q^2+p^2r^2)^2} + \frac{B_0(B_0-p)}{q^2+p^2r^2} + A_0^2p^2r^2$$

There are infinitely many bound energy levels clustered into bands of two states and tunneling of states between the left and right wells. See Fig. 5.

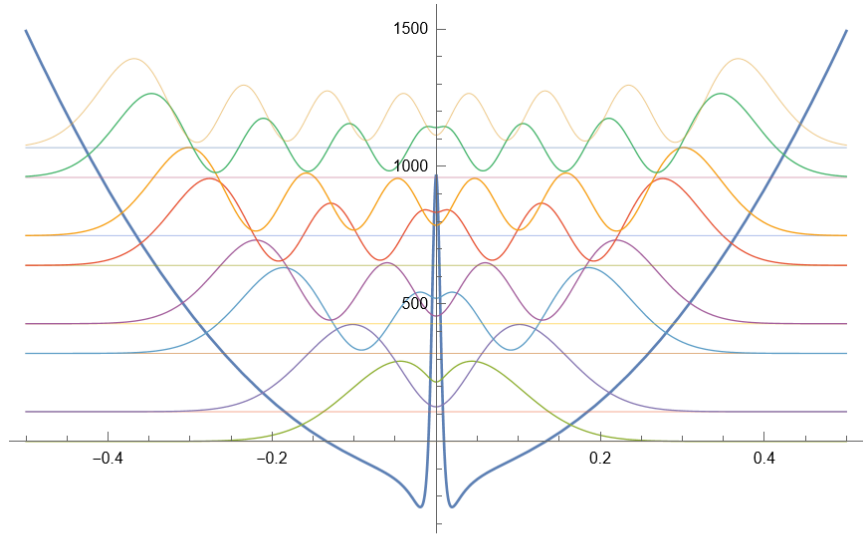

Fig 5. Potential G301t2. Complexification of the 3D-Oscillator with tunneling,

for A0=100, B0=0.13, p=0.8, q=0.009

Case 3: The sequence $a_k = (A_k, B_k)$ is given by
$$A_k = A_0, \ B_k = B_0 + kp$$

The energy values are given by
$$E_n = 4A_0pn$$

The real part of the complex potential is the same as above.
There are infinitely many bound energy levels and tunneling of states between the left and right wells.

## 4.4 Group 4: Superpotential with Exponential function and a constant (Complexification of Morse potential)

### 4.4.1 Superpotential $W(x, A_0, B_0) = A_0 - B_0 e^{-(px+iq)}$

The sequence $\{a_k = (A_k, B_k)\}$ is given by
$$A_k = A_0 - kp, \quad B_k = B_0$$

The energy values are given by
$$E_n = A_0^2 - A_n^2$$

The real part of the complex potential is
$$V(x) = A_0^2 + B_0 e^{-2px}(-2pxe^{px}(2A_0 + p)\cos q + B_0 \cos 2q)$$

This is the complexification of Morse's potential. There is a finite number of bound states, their numbers do not exceed $\frac{2A_0+p}{2p}$ with $0 < p < 2A_0$.

We will see in a future paper that this superpotential is a special case of a superpotential obtained as a combination of hyperbolic functions and a constant.

## 4.5 Group 5: Superpotentials with one trigonometric term coth and a constant.

### 4.5.1 Superpotential $W(x, A_0, B_0) = rA_0 \coth r(px + iq) - B_0$

This potential was obtained by introducing the ansatz superpotential
$$W(x, A_0, B_0) = rA_0 \coth r(px + iq) - B_0, \tag{46}$$
an extension to the Eckart potential ($r = 1, q = 0$), with the partner potentials satisfying the shape invariance condition (19). The parameters, $p, q,$ and $r$ are fixed positive real constants. For practical reasons, I will assume all coefficients $A_0$, and $B_0$ to be positive. For the other cases, we only need to change the sign of the coefficient in question and perform a new analysis. The complex potential $V(x) = V_0(x, A_0, B_0)$ of the second-order differential equation (1) and the Schrödinger equation (11) is given by

$$V_0(x, A_0, B_0) = W^2(x, A_0, B_0) - W'(x, A_0, B_0) \tag{47}$$

I will consider its real part for its physical meaning, the real part of the potential of $V(x)$ is then given by

$$V_-(x, A_0, B_0) \tag{48}$$
$$= \frac{2B_0^2 - 2A_0(A_0+2p)r^2 + (B_0^2 + A_0^2 r^2)\cos 4qr + (B_0^2 + A_0^2 r^2)\cosh 4prx}{2(\cos 2qr - \cosh 2prx)^2} +$$

$$\frac{4\cos 2qr\left((-B_0^2+pA_0r^2)\cosh 2prx+A_0B_0r\sinh 2prx\right)-2A_0B_0r\sinh 4prx}{2(\cos 2qr-\cosh 2prx)^2}$$

As we can see, the expression of the potential above is quite long to describe in writing, therefore, for future potentials, I will only give the superpotential $W$ and leave it to the reader to use their preferred CAS software to extract the real part of the potential from (47).

It can be seen that the value $Re(V)_{min} = B_0^2 + A_0r^2 - (A_0+p)r^2 csch^2(qr)$ is the minimum attained at $x=0$. This minimum cannot exceed $B_0^2 - pr^2$.

The sequence $a_k = (A_k, B_k)$ in the definition of the shape invariance property is defined as follows:
$$A_k = A_0 - kp, \tag{49}$$
$$B_k = \frac{A_0 B_0}{A_0 - kp} \quad \text{for } k = 0,1,2,.. \tag{50}$$

The first energy level (eigenvalue) of the potential $V$ is,
$$E_1^{(-)} = V_+(x, A_0, B_0) - V_-(x, A_1, B_1) = \frac{p(2A_0-p)(-B_0^2+(A_0-p)^2r^2)}{(A_0-p)^2} \tag{51}$$
with the condiion that $E_1 = E_1^{(-)}$ is positive (18)

$$E_1 = r^2 A_0^2 + B_0^2 - (r^2 A_1^2 + B_1^2), \tag{52}$$

Thus, for all $n = 0,1,2,...$, the $n^{th}$ bound energy level is given by
$$E_n = r^2 A_0^2 + B_0^2 - (r^2 A_n^2 + B_n^2) \tag{53}$$
$$= B_0^2 + r^2 A_0^2 - \frac{A_0^2 B_0^2}{(A_0-np)^2} - r^2(A_0-np)^2 \tag{54}$$

### 4.5.2 Superpotential $W(x, A_0, B_0) = rA_0 \coth r(px+iq) - iB_0$

The sequence $a_k = (A_k, B_k)$ is defined as follows:
$$A_k = A_0 - kp, \tag{55}$$
$$B_k = \frac{A_0 B_0}{A_0 - kp} \quad \text{for } k = 0,1,2,.. \tag{56}$$

the $n^{th}$ bound energy level is given by
$$E_n = r^2 A_0^2 - B_0^2 - (r^2 A_n^2 - B_n^2) \tag{57}$$
$$= -B_0^2 + r^2 A_0^2 + \frac{A_0^2 B_0^2}{(A_0-np)^2} - r^2(A_0-np)^2 \tag{58}$$

To have an ordered ascendingly eigenvalues $E_n$, the coefficients should satisfy the following:

$B_0 > 0$, $p > 0$, $r > 0$, $\frac{2p}{3} < A_0 \leq 2p$, and the number of bound states $1 \leq n < \frac{2A_0+p}{2p}$, or $B_0 > 0$, $p > 0$, $r > 0$, $A_0 > 2p$ and $1 \leq n < \frac{A_0}{p}$.

Two BICs can be seen in Fig. 6 if the parameters are chosen to be; $A_0 = 7.81$, $B_0 = 0.2$, $p = 1.2$, $r = 0.3$, and $q = 6.1$.

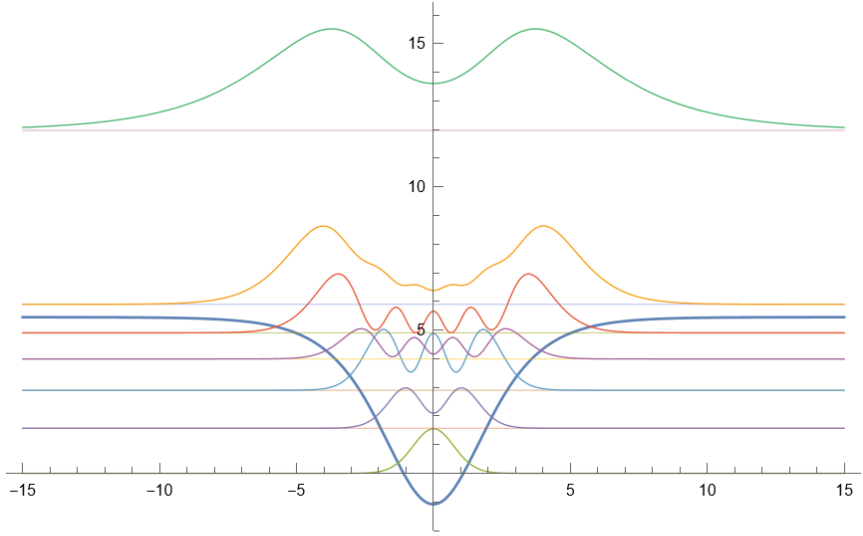

Fig 6. Potential G502. Complexification of the extended Ekhart potential with 2 BIC's, parameters A0=7.81,B0=0.2,p=1.2,r=0.3,q=6.1

### 4.5.3   Superpotential $W(x, A_0, B_0) = -rA_0 \cot r(px + iq) + B_0$

The sequence $a_k = (A_k, B_k)$ is defined as follows:
$$A_k = A_0 + kp, \tag{59}$$
$$B_k = \frac{A_0 B_0}{A_0 + kp} \quad \text{for } k = 0,1,2,.. \tag{60}$$

The $n^{th}$ bound energy level is given by
$$E_n = -r^2 A_0^2 + B_0^2 - (r^2 A_n^2 + B_n^2) \tag{61}$$
$$= B_0^2 - r^2 A_0^2 - \frac{A_0^2 B_0^2}{(A_0+np)^2} + r^2 (A_0 + np)^2 \tag{62}$$

There is an infinite number of bound states.
When $q = 0$ and $r = 1$, we recover the Rosen-Morse I potential.

## 4.6   Group 6: Superpotentials with one trigonometric term cot and one constant.

### 4.6.1   Superpotential $W(x, A_0, B_0) = -rA_0 \cot r(px + iq) + B_0$

The sequence $a_k = (A_k, B_k)$ is defined as follows:
$$A_k = A_0 + kp, \tag{63}$$
$$B_k = \frac{A_0 B_0}{A_0 + kp} \quad \text{for } k = 0,1,2,.. \tag{64}$$

the $n^{th}$ bound energy level is given by
$$E_n = -r^2 A_0^2 + B_0^2 - (-r^2 A_n^2 + B_n^2) \tag{65}$$
$$= B_0^2 - r^2 A_0^2 - \frac{A_0^2 B_0^2}{(A_0+np)^2} + r^2(A_0 + np)^2 \tag{66}$$

This is the complexification of the Rosen-Morse I potential which is a periodic potential.

### 4.6.2  Superpotential $W(x, A_0, B_0) = -rA_0 \cot r(px + iq) + iB_0$

The sequence $a_k = (A_k, B_k)$ is defined as follows:
$$A_k = A_0 + kp, \tag{67}$$
$$B_k = \frac{A_0 B_0}{A_0 + kp} \quad \text{for } k = 0,1,2,.. \tag{68}$$

the $n^{th}$ bound energy level is given by
$$E_n = -r^2 A_0^2 - B_0^2 - (-r^2 A_n^2 - B_n^2) \tag{69}$$
$$= -B_0^2 - r^2 A_0^2 + \frac{A_0^2 B_0^2}{(A_0+np)^2} + r^2(A_0 + np)^2 \tag{70}$$

if $0 < B_0 < r(A_0 + p)$, we have an infinite number of bound states. This is a symmetric periodic potential, the extension of the Rosen-Morse I potential. seen in Fig. 7

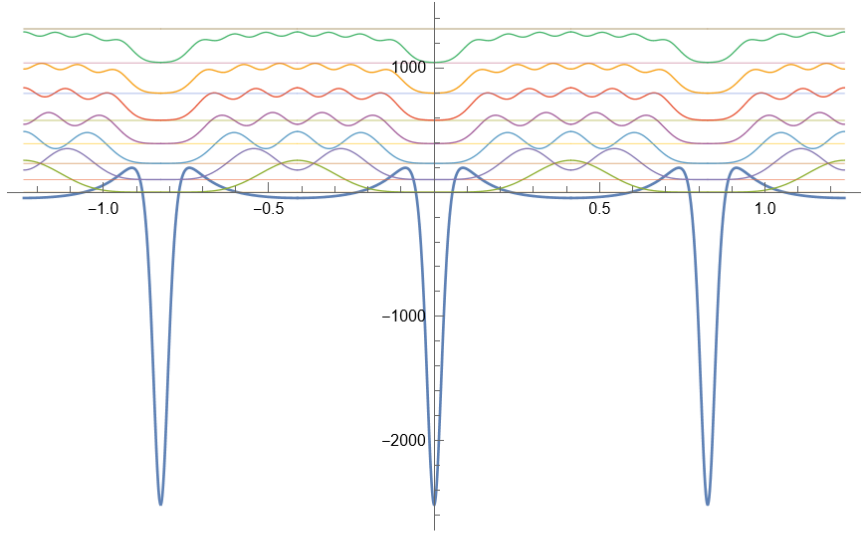

Fig 7. Potential G601. Complexification of the extended periodic Rosen-Morse I potential, parameters A0=6,B0=0.1,p=2,q=0.1,r=1.9

## 4.7  Group 7: Superpotentials with one trigonometric term tanh and one constant.

### 4.7.1 Superpotential $W(x, A_0, B_0) = rA_0\tanh r(px + iq) - B_0$

The sequence $a_k = (A_k, B_k)$ is defined as follows:
$$A_k = A_0 - kp, \tag{71}$$
$$B_k = \frac{A_0 B_0}{A_0 - kp} \quad \text{for } k = 0, 1, 2, \ldots \tag{72}$$

The $n^{th}$ bound energy level is given by
$$E_n = r^2 A_0^2 + B_0^2 - (r^2 A_n^2 + B_n^2) \tag{73}$$
$$= B_0^2 + r^2 A_0^2 - \frac{A_0^2 B_0^2}{(A_0 - np)^2} - r^2(A_0 - np)^2 \tag{74}$$

To have an ordered ascendingly eigenvalues $E_n$, and the normalizability of the eigenfunctions, the coefficients should satisfy the following:

$$0 < A0 < p, \; B_0 > 0, \; 0 < r < -\frac{A_0 B_9}{(A_0 - p)p} \text{ and } 1 \leq n < \frac{A_0}{2p} + \frac{1}{2p}\sqrt{\frac{(4A_0 B_0 + A_0^2 r)}{r}}.$$

When $q = 0$ and $r = 1$ we obtain Rosen-Morse II potential as a special case.

For a general case when $r$ is an arbitrary nonzero real number, different from the special cases above, I obtain the following:

### 4.7.2 Superpotential $W(x, A_0, B_0) = rA_0\tanh r(px + iq) - iB_0$

The sequence $a_k = (A_k, B_k)$ is defined as follows:
$$A_k = A_0 - kp, \tag{75}$$
$$B_k = \frac{A_0 B_0}{A_0 - kp} \quad \text{for } k = 0, 1, 2, \ldots \tag{76}$$

The $n^{th}$ bound energy level is given by
$$E_n = r^2 A_0^2 - B_0^2 - (r^2 A_n^2 - B_n^2) \tag{77}$$
$$= -B_0^2 + r^2 A_0^2 + \frac{A_0^2 B_0^2}{(A_0 - np)^2} - r^2(A_0 - np)^2 \tag{78}$$

To have an ordered ascendingly eigenvalues $E_n$, and the normalizability of the eigenfunctions, the coefficients should satisfy the following:

$$0 < A_0 < p, \; B_0 > 0, 0 < r < -\frac{A_0 B_9}{(A_0 - p)p} \text{ and } 1 \leq n < \frac{A_0}{2p} + \frac{1}{2p}\sqrt{\frac{(4A_0 B_0 + A_0^2 r)}{r}}$$

This potential possesses BICs. The complexification of the complexified Rosen-Morse II was introduced in [3]. See Fig. 8.

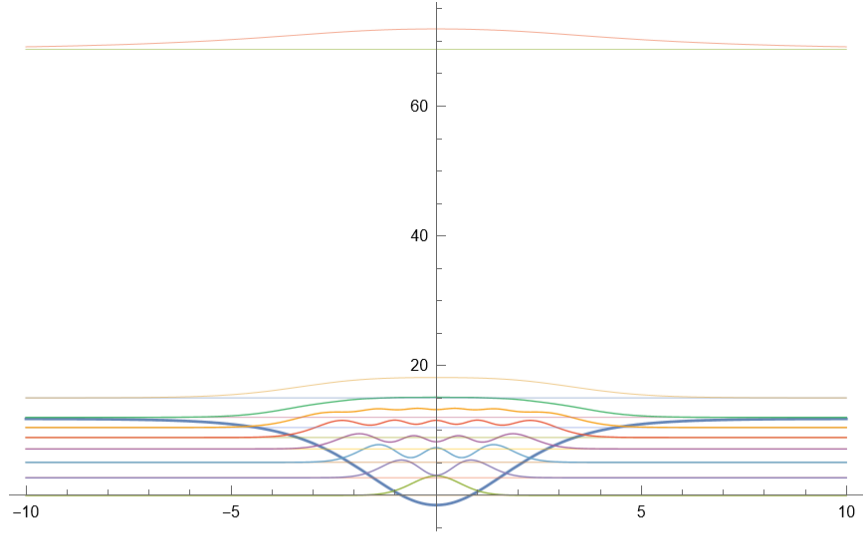

Fig 8. Potential G702, Complexification of the extended Rosen-Morse II potential

with two BIC's, parameters A0=7.498,C0=0.3,p=0.9,q=0.03,r=0.46

## 4.8 Group 8: Superpotentials with one trigonometric term tan and one constant.

### 4.8.1 Superpotential $W(x, A_0, B_0) = rA_0 \tan r(px + iq) + B_0$

The sequence $a_k = (A_k, B_k)$ is defined as follows:
$$A_k = A_0 + kp, \tag{79}$$
$$B_k = \frac{A_0 B_0}{A_0 + kp} \quad \text{for } k = 0,1,2,.. \tag{80}$$

The $n^{th}$ bound energy level is given by
$$E_n = -r^2 A_0^2 + B_0^2 - (-r^2 A_n^2 + B_n^2) \tag{81}$$
$$= B_0^2 - r^2 A_0^2 - \frac{A_0^2 B_0^2}{(A_0 + np)^2} - r^2(A_0 + np)^2 \tag{82}$$

an infinite number of bound states.
Next, the complexified superpotential constant.$B_0$.

### 4.8.2 Superpotential $W(x, A_0, B_0) = rA_0 \tan r(px + iq) + iB_0$

The sequence $a_k = (A_k, B_k)$ is defined as follows:
$$A_k = A_0 + kp,$$

$$B_k = \frac{A_0 B_0}{A_0 + kp} \text{ for } k = 0,1,2,..$$

The $n^{th}$ bound energy level is given by
$$E_n = -r^2 A_0^2 - B_0^2 + (r^2 A_n^2 + B_n^2) \tag{83}$$
$$= -B_0^2 - r^2 A_0^2 + \frac{A_0^2 B_0^2}{(A_0 + np)^2} + r^2 (A_0 + np)^2 \tag{84}$$

Infinite number of bound states when $A_0 > 0, B_0 > 0, p > 0, r > 0, \frac{B_0}{A_0} \leq r$.

## 4.9 Group 9: Superpotentials with two trigonometric terms tan and cot.

### 4.9.1 Superpotential $W(x, A_0, B_0) = A_0 \tan(px + iq) - B_0 \cot(px + iq)$

This is a completely new potential. I have never seen one before. The description is given in three cases:

case 1: The sequence $a_k = (A_k, B_k)$ is defined as follows:

$$A_k = (-1)^k A_0, \tag{85}$$
$$B_k = B_0 + kp \text{ for } k = 0,1,2,.. \tag{86}$$

The $n^{th}$ bound energy level is given by
$$E_n = (A_n + B_n)^2 - (A_0 + B_0)^2 \tag{87}$$
$$= -(A_0 + B_0)^2 + ((-1)^n A_0 + B_0 + np)^2 \tag{88}$$

Infinite number of bound states when $B_0 > 0$, $p > 0$, and $0 < A_0 < \frac{p}{2}$.

Fig. 9 for case 1.

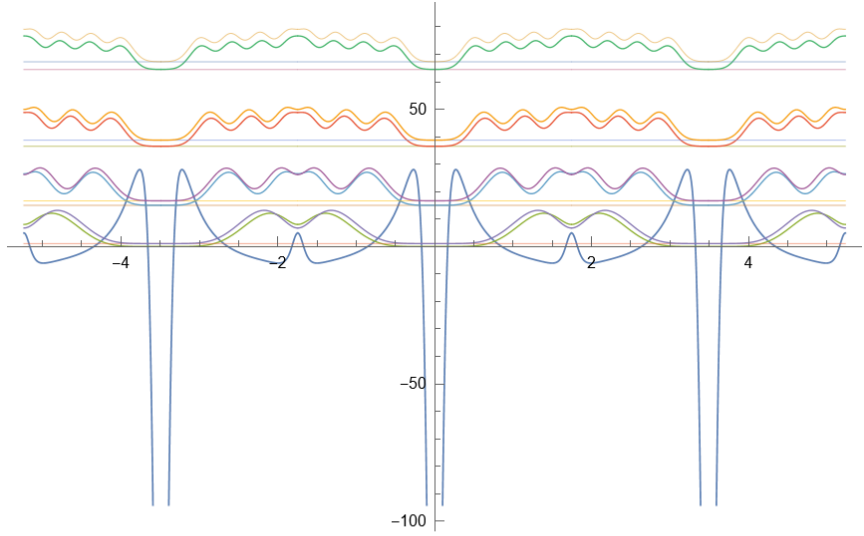

Fig 9. Potential G901, a new periodic potential with tunneling,

A0=0.37,B0=2.9,p=0.9,q=0.14

case 2: The sequence $a_k = (A_k, B_k)$ is defined as follows:
$$A_k = A_0 + kp, \tag{89}$$
$$B_k = (-1)^k B_0 \text{ for } k = 0,1,2,.. \tag{90}$$

The $n^{th}$ bound energy level is given by
$$E_n = (A_n + B_n)^2 - (A_0 + B_0)^2 \tag{91}$$
$$= -(A_0 + B_0)^2 + (A_0 + (-1)^n B_0 + np)^2 \tag{92}$$

Infinite number of bound states when $A_0 > 0, p > 0,$ and $0 < B_0 < \frac{p}{2}$.

case 3: The sequence $a_k = (A_k, B_k)$ is defined as follows:
$$A_k = A_0 + kp, \tag{93}$$
$$B_k = B_0 + kp \text{ for } k = 0,1,2,.. \tag{94}$$

the $n^{th}$ bound energy level is given by
$$E_n = (A_n + B_n)^2 - (A_0 + B_0)^2 \tag{95}$$
$$= -(A_0 + B_0)^2 + (A_0 + B_0 + 2np)^2 \tag{96}$$

An infinite number of bound states when $A_0 > 0, B_0 > 0, p > 0$.
When $q = 0,$ we find the Pöschl-Teller I potential in the last case.

### 4.9.2  Superpotential $W(x, A_0, B_0) = A_0 \tan(px + iq) - 2B_0 \cot 2(px + iq)$

Three cases:
*case 1*: The sequence $a_k = (A_k, B_k)$ is defined as follows:

$$A_k = A_0 \tag{97}$$
$$B_k = B_0 + kp \quad \text{for } k = 0,1,2,\ldots \tag{98}$$

The $n^{th}$ bound energy level is given by
$$E_n = (A_n + 2B_n)^2 - (A_0 + 2B_0)^2 \tag{99}$$
$$= -(A_0 + 2B_0)^2 + (A_0 + 2B_0 + 2np)^2 \tag{100}$$

An infinite number of bound states when $0 < A_0 < \frac{1}{2}(-2B_0 + p),\ 0 < B_0 < \frac{p}{2}$.

This case produces periodic potentials.

case 2: The sequence $a_k = (A_k, B_k)$ is defined as follows:
$$A_k = \begin{cases} -A_0 - 2B_0 - kp, & k \text{ is odd} \\ A_0 - kp, & k \text{ is even} \end{cases} \tag{101}$$
$$B_k = B_0 + kp \quad \text{for } k = 0,1,2,\ldots \tag{102}$$

The $n^{th}$ bound energy level when $n$ is even is given by
$$E_n = (A_n + 2B_n)^2 - (A_0 + 2B_0)^2 \tag{103}$$
$$= -(A_0 + 2B_0)^2 + (A_0 + 2B_0 + np)^2 \tag{104}$$

and when $n$ is odd
$$E_n = -(A_0 + 2B_0)^2 + (-A_0 + np)^2 \tag{105}$$

An infinite number of bound states when $p > 0,\ 0 < A_0 < \frac{1}{2}(-2B_0 + p)$ and $0 < B_0 < \frac{p}{2}$.

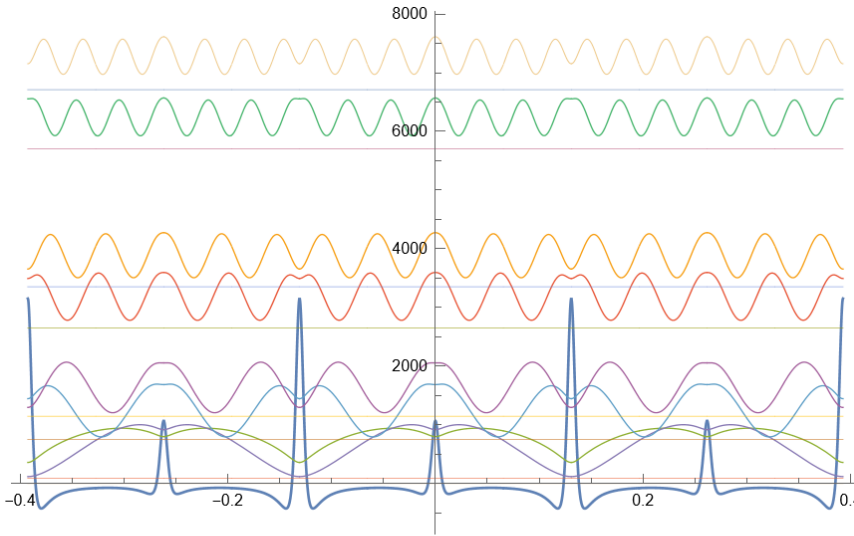

Fig 10. Periodic potential G902 and the only allowed first seven states with tunneling of lower states and 3 BIC's, A0=2, B0=0.8, p=12, q=0.09

case 3: The sequence $a_k = (A_k, B_k)$ is defined as follows:
$$A_k = \begin{cases} A_0 + 2B_0 + kp, & k \text{ is odd} \\ A_0 + kp, & k \text{ is even} \end{cases} \tag{106}$$

$$B_k = (-1)^k B_0 \quad \text{for } k = 0,1,2,..  \tag{107}$$

the $n^{th}$ bound energy level is given by for even bound states
$$E_n = (A_n + 2B_n)^2 - (A_0 + 2B_0)^2 \tag{108}$$
$$= -(A_0 + 2B_0)^2 + (A_0 + 2B_0 + np)^2 \tag{109}$$

for odd states
$$E_n = (A_n + 2B_n)^2 - (A_0 + 2B_0)^2 \tag{110}$$
$$= -(A_0 + 2B_0)^2 + (A_0 + np)^2 \tag{111}$$

An infinite number of bound states when $A_0 > 0$, $B_0 > 0$, $p > 0$.
In general, when $r$ is taken to be an arbitrary number $r > 1$, a formula can be found.

### 4.9.3  Superpotential $W(x, A_0, B_0) = A_0 \tan(px + iq) - rB_0 \cot r(px + iq)$

The sequence $a_k = (A_k, B_k)$ is defined as follows:

$$A_k = A_0 + kp \tag{112}$$
$$B_k = B_0 + kp \quad \text{for } k = 0,1,2,.. \tag{113}$$

The $n^{th}$ bound energy level is given by
$$E_n = (A_n + rB_n)^2 - (A_0 + rB_0)^2 \tag{114}$$
$$= -(A_0 + rB_0)^2 + (A_0 + rB_0 + (1+r)np)^2 \tag{115}$$

A more general case can be derived for any $r$ and $s$ positive real numbers:

### 4.9.4  Superpotential $W(x, A_0, B_0) = sA_0 \tan s(px + iq) - rB_0 \cot r(px + iq)$

The sequence $a_k = (A_k, B_k)$ is defined as follows:

$$A_k = A_0 + kp \tag{116}$$
$$B_k = B_0 + kp \quad \text{for } k = 0,1,2,.. \tag{117}$$

The $n^{th}$ bound energy level is given by
$$E_n = (sA_n + rB_n)^2 - (sA_0 + rB_0)^2 \tag{118}$$
$$= -(sA_0 + rB_0)^2 + (sA_0 + rB_0 + (s+r)np)^2 \tag{119}$$

An infinite number of bound states in ascending order when $0 < s < r$, $p > 0$, and $B_0 > 0$.

## 4.10   Group 10: Superpotentials with two hyperbolic terms tanh and coth.

### 4.10.1 Superpotential $W(x, A_0, B_0) = A_0 \tanh(px + iq) + B_0 \coth(px + iq)$

Case1:

$$A_k = (-1)^k A_0 \qquad (120)$$
$$B_k = B_0 - kp \quad \text{for } k = 0,1,2,.. \qquad (121)$$

The $n^{th}$ bound energy level is given by
$$E_n = (A_0 + B_0)^2 - (A_n + B_n)^2 \qquad (122)$$
$$= -(A_0 + B_0)^2 - ((-1)^n A_0 + B_0 - np)^2 \qquad (123)$$

$$A_0 < \frac{p}{2}, B_0 > \frac{5p}{2}, \qquad 1 \leq n < \frac{2B_0 - p}{4p}.$$

Case2:

$$A_k = A_0 - kp \qquad (124)$$
$$B_k = (-1)^k B_0 \quad \text{for } k = 0,1,2,.. \qquad (125)$$

The $n^{th}$ bound energy level is given by
$$E_n = (A_0 + B_0)^2 - (A_n + B_n)^2 \qquad (126)$$
$$= -(A_0 + B_0)^2 - (A_0 + (-1)^n B_0 - 2np)^2 \qquad (127)$$

$$A_0 > \frac{5p}{2}, B_0 < \frac{p}{2}, \qquad 1 \leq n < \frac{2A_0 - p}{4p}.$$

Case3:

$$A_k = A_0 - kp \qquad (128)$$
$$B_k = B_0 - kp \quad \text{for } k = 0,1,2,.. \qquad (129)$$

The $n^{th}$ bound energy level is given by
$$E_n = (A_0 + B_0)^2 - (A_n + B_n)^2 \qquad (130)$$
$$= -(A_0 + B_0)^2 - (A_0 + B_0 - 2np)^2 \qquad (131)$$

$(0 < B_0 \leq p, \ A_0 > -B_0 + p, \ 1 \leq n < \frac{A_0 + B_0 + p}{2p})$ or

$$(B_0 > p, \qquad A_0 > 0, \qquad 1 \leq n < \frac{A_0 + B_0 + p}{2p}).$$

When we set $q = 0$ in this last case, we find the Pöschl-Teller II potential.

### 4.10.2 Superpotential $W(x, A_0, B_0) = A_0 \tanh(px + iq) + 2B_0 \coth 2(px + iq)$

Case1:
$$A_k = A_0 \tag{132}$$
$$B_k = B_0 - kp \quad \text{for } k = 0,1,2,.. \tag{133}$$

The $n^{th}$ bound energy level is given by
$$E_n = (A_0 + 2B_0)^2 - (A_n + 2B_n)^2 \tag{134}$$
$$= -(A_0 + 2B_0)^2 - (A_0 + 2B_0 - 2np)^2 \tag{135}$$

with finite bound states when $\quad 0 < p < A_0 + 2B_0, \ 1 \le n < \frac{A_0 + 2B_0 - p}{2p}$.

Case2:

$$A_k = \begin{cases} A_0 + 2B_0 - kp, k \text{ is odd} \\ A_0 - kp, k \text{ is even} \end{cases} \tag{136}$$
$$B_k = (-1)^k B_0 \quad \text{for } k = 0,1,2,.. \tag{137}$$

The $n^{th}$ bound energy level is given by
$$E_n = (A_0 + 2B_0)^2 - (A_n + 2B_n)^2 \tag{138}$$
$$= \begin{cases} (A_0 + 2B_0)^2 - (A_0 - np)^2, if\ n\ is\ odd \\ (A_0 + 2B_0)^2 - (A_0 + 2B_0 - np)^2, if\ n\ is\ even \end{cases} \tag{139}$$

The condition to have increasing eigenvalues with at least three excited states is $\varepsilon(2k + 1) > \varepsilon(2k) > \varepsilon(2k - 1) > 0$, which requires the following restrictions on the parameters and $k$
$$0 < B_0 < \frac{p}{2}, A_0 > \frac{1}{2}(-2B_0 + 5p), \qquad 1 \le k < \frac{2A_0 + 2B_0 - p}{4p}.$$
See Fig. 11.

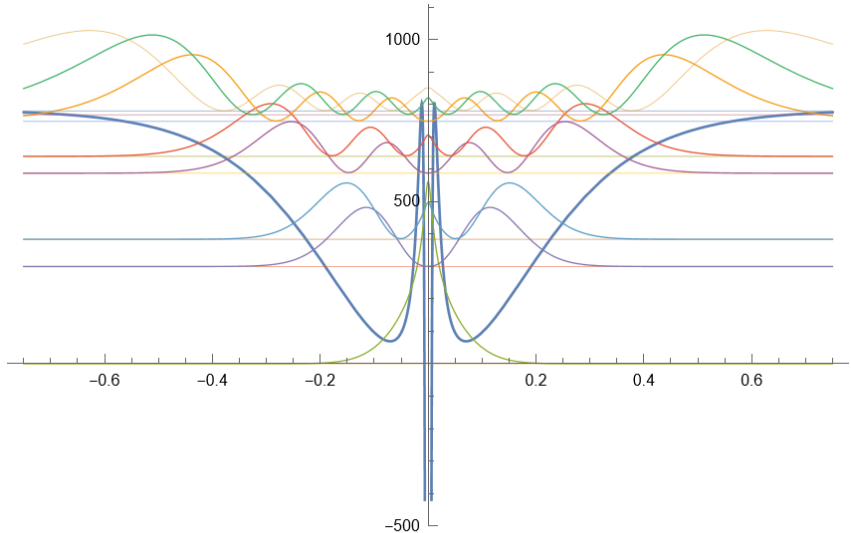

Fig 11. Potential G1002, tunneling and band structure, A0=26,B0=1,p=4,q=0.027

Case3:

$$A_k = \begin{cases} -A_0 - 2B_0 + kp, k \text{ is odd} \\ A_0 + kp, k \text{ is even} \end{cases} \quad (140)$$

$$B_k = (-1)^k B_0 \quad \text{for } k = 0,1,2,.. \quad (141)$$

The $n^{th}$ bound energy level is given by
$$E_n = (A_0 + 2B_0)^2 - (A_n + 2B_n)^2 \quad (142)$$
$$= \begin{cases} (A_0 + 2B_0)^2 - (-A_0 - 4B_0 + np)^2, \text{if } n \text{ is odd} \\ (A_0 + 2B_0)^2 - (A_0 + 2B_0 + np)^2, \text{if } n \text{ is even} \end{cases} \quad (143)$$

when $\frac{p}{2} < B_0 < \frac{3p}{2}$, $A_0 > 0$, in this case there are only two excited states and we can observe tunneling of the second excited states between the three adjacent wells. Plot not shown here.

### 4.10.3 Superpotential $W(x, A_0, B_0) = A_0 \tanh(px + iq) + rB_0 \coth r(px + iq)$

$$A_k = A_0 - kp \quad (144)$$
$$B_k = B_0 - kp \quad \text{for } k = 0,1,2,.. \quad (145)$$

The $n^{th}$ bound energy level is given by
$$E_n = (A_0 + rB_0)^2 - (A_n + rB_n)^2 \quad (146)$$
$$= -(A_0 + 2B_0)^2 - (A_0 + rB_0 - (1+r)np)^2 \quad (147)$$

with finite bound states when $0 < r < 1, 0 < p > 2B_0$, and the number of excited states is such that $1 \leq n < \frac{-2B_0 + 3p + 2B_0 r + pr}{2p(1+r)}$, or $r > 1, 0 < p < 2B_0$ and $1 \leq n < \frac{-2B_0 + 3p + 2B_0 r + pr}{2p(1+r)}$.

Let's look at a more general case.

### 4.10.4 Superpotential $W(x, A_0, B_0) = sA_0 \tanh s(px + iq) + rB_0 \coth r(px + iq)$

$$A_k = A_0 - kp \quad (148)$$
$$B_k = B_0 - kp \quad \text{for } k = 0,1,2,.. \quad (149)$$

The $n^{th}$ bound energy level is given by
$$E_n = (sA_0 + rB_0)^2 - (sA_n + rB_n)^2 \quad (150)$$
$$= -(sA_0 + rB_0)^2 - (sA_0 + rB_0 - (s+r)np)^2 \quad (151)$$

with finite bound states when $r > s > 0, B_0 > \frac{p(3r+s)}{2(r-s)}$, $1 \leq n < \frac{2B_0 r - p - 2B_0 s + pr}{2p(s+r)}$.

## 4.11 Group 11. Superpotential with two hyperbolic terms tan and sec

### 4.11.1 Superpotential $W(x, A_0, B_0) = A_0 \tan(px + iq) + B_0 \sec(px + iq)$

We have three cases,
Case1: The squeeze $\{a_k\}$ is defined by

$$A_k = \begin{cases} -B_0 + \frac{kp}{2}, k \text{ is odd} \\ A_0 + \frac{kp}{2}, k \text{ is even} \end{cases} \quad (152)$$

$$B_k = \begin{cases} -A_0 + \frac{kp}{2}, k \text{ is odd} \\ B_0 + \frac{kp}{2}, k \text{ is even} \end{cases} \text{ for } k = 0,1,2,.. \quad (153)$$

The eigenvalues are given by
$$E_n = -A_0^2 + A_n^2 \quad (154)$$

The ordering condition on the eigenvalues will require $0 < B_0 < \frac{p}{2}$, $0 < A_0 < \frac{1}{2}(-2B_0 + p)$, and an infinite number of bound states with discrete states in the continuum.

This potential is periodic and can possess an infinite number of wells with tunneling through the forbidden regions happening with high probabilities. See Fig. 12.

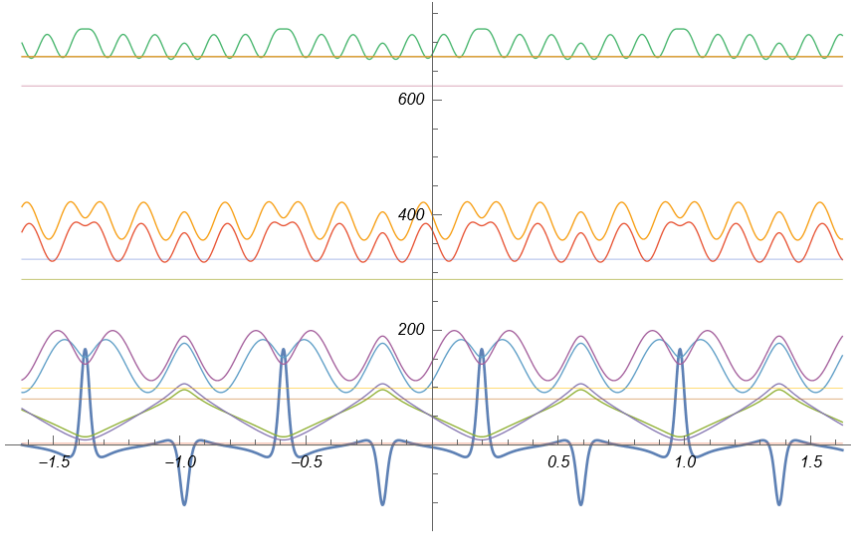

Fig 12. Potential G1201 with tunneling and resonant states for A0=1,B0=2, p=8,q=0.3

Case2.
The sequence $\{a_k\}$ is defined by

$$A_k = \begin{cases} B_0 + \frac{kp}{2}, k \text{ is odd} \\ A_0 + \frac{kp}{2}, k \text{ is even} \end{cases} \quad (155)$$

$$B_k = \begin{cases} A_0 + \frac{kp}{2}, k \text{ is odd} \\ B_0 + \frac{kp}{2}, k \text{ is even} \end{cases} \text{ for } k = 0,1,2,..\qquad(156)$$

The eigenvalues are given by
$$E_n = -A_0^2 + A_n^2 \qquad(157)$$

The ordering condition on the eigenvalues will require $0 < B_0 \leq \frac{p}{2}$, $0 < A_0 < \frac{1}{2}(2B_0 + p)$ or $B_0 > \frac{p}{2}$, $\frac{1}{2}(2B_0 - p) < A_0 < \frac{1}{2}(2B_0 + p)$ with an infinite number of bound states.

The same observation as before, this potential is periodic and can possess an infinite number of wells with tunneling through the forbidden regions happening with high probabilities.

Case3. The sequence $a_k = (A_k, B_k)$ is defined as follows:
$$A_k = A_0 + kp, \qquad(158)$$
$$B_k = B_0 \text{ for } k = 0,1,2,.. \qquad(159)$$

The eigenvalues are given by
$$E_n = -A_0^2 + A_n^2 \qquad(160)$$

This is the complexification of the trigonometric Scarf I potential. It can be observed that Scarf I potential has all these hidden properties that were missing in the real case. A periodic potential with multi-wells, plenty of tunneling between adjacent wells, resonant states, and BIC states.

## 4.12 Group 12. Superpotential with two hyperbolic terms tanh and sech

### 4.12.1 Superpotential $W(x, A_0, B_0) = A_0 \tanh(px + iq) + B_0 sech(px + iq)$

The sequence $a_k = (A_k, B_k)$ is defined as follows:
$$A_k = A_0 - kp, \qquad(161)$$
$$B_k = B_0 \text{ for } k = 0,1,2,.. \qquad(162)$$

The eigenvalues are given by
$$E_n = A_0^2 - A_n^2 \qquad(163)$$

This is the complexification of the hyperbolic Scarf II potential, see Fig.13. The maximum and minimum of the potential are not shown.

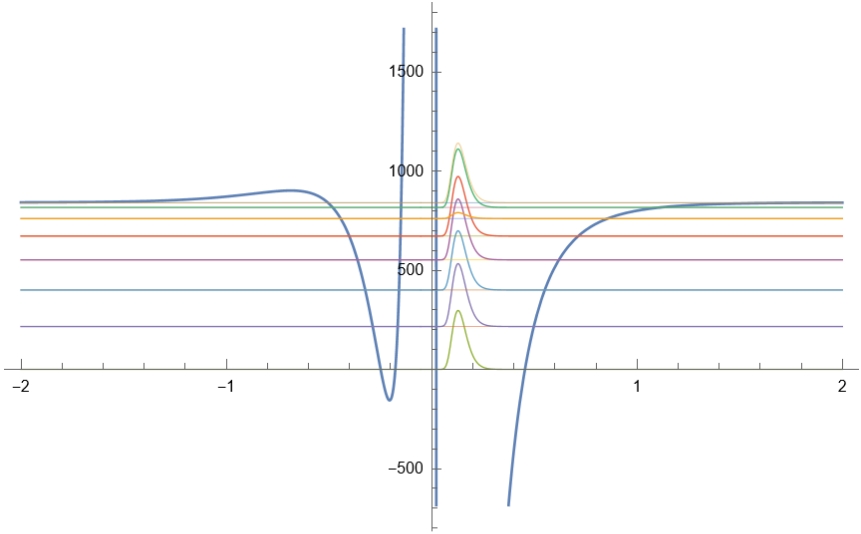

Fig 13. Potential G1200 with a very deep double well and the only allowed

seven energy levels(n<A0/p) for A0=29,B0=74,p=4,q=4.5

## 4.13 Group 13. Superpotential with two hyperbolic terms coth and csch

### 4.13.1 Superpotential $W(x, A_0, B_0) = A_0\coth(px + iq) + B_0 csch(px + iq)$

Three cases:
Case 1: The sequence $\{a_k\}$ is defined by

$$A_k = \begin{cases} B_0 - \frac{kp}{2}, k \text{ is odd} \\ A_0 - \frac{kp}{2}, k \text{ is even} \end{cases} \quad (164)$$

$$B_k = \begin{cases} A_0 - \frac{kp}{2}, k \text{ is odd} \\ B_0 - \frac{kp}{2}, k \text{ is even} \end{cases} \text{ for } k = 0,1,2,.. \quad (165)$$

The eigenvalues are given by
$$E_n = A_0^2 - A_n^2 \quad (166)$$

The ordering condition on the eigenvalues will require $p < B_0 \leq \frac{3p}{2}$, $\frac{1}{2}(-2B_0 + 5p) < A_0 < \frac{1}{2}(2B_0 + p)$ or $B_0 > \frac{3p}{2}$, $\frac{1}{2}(2B_0 - p) < A_0 < \frac{1}{2}(2B_0 + p)$ with $n = 2k - 1$, the number of bound states is satisfying the condition $1 \leq k < \frac{2A_0 + 2B_0 - p}{4p}$.

This potential has double wells which can exhibit tunneling through the central forbidden region.
Case 2: The sequence $\{a_k\}$ is defined by

$$A_k = \begin{cases} -B_0 - \frac{kp}{2}, k \text{ is odd} \\ A_0 - \frac{kp}{2}, k \text{ is even} \end{cases} \quad (167)$$

$$B_k = \begin{cases} A_0 + \frac{kp}{2}, k \text{ is odd} \\ B_0 + \frac{kp}{2}, k \text{ is even} \end{cases} \text{ for } k = 0,1,2,.. \quad (168)$$

The eigenvalues are given by
$$E_n = A_0^2 - A_n^2 \quad (169)$$

This potential has a maximum of two excited states, or a three-state system, under the condition $\frac{1}{2}(2B_0 + p) < A_0 < \frac{1}{2}(2B_0 + 3p)$.

Case 3. The sequence $a_k = (A_k, B_k)$ is defined as follows:
$$A_k = A_0 - kp, \quad (170)$$
$$B_k = B_0 \text{ for } k = 0,1,2,.. \quad (171)$$

The eigenvalues are given by
$$E_n = A_0^2 - A_n^2 \quad (172)$$

This is the complexification of Pöschl-Teller I potential, Fig. 14.

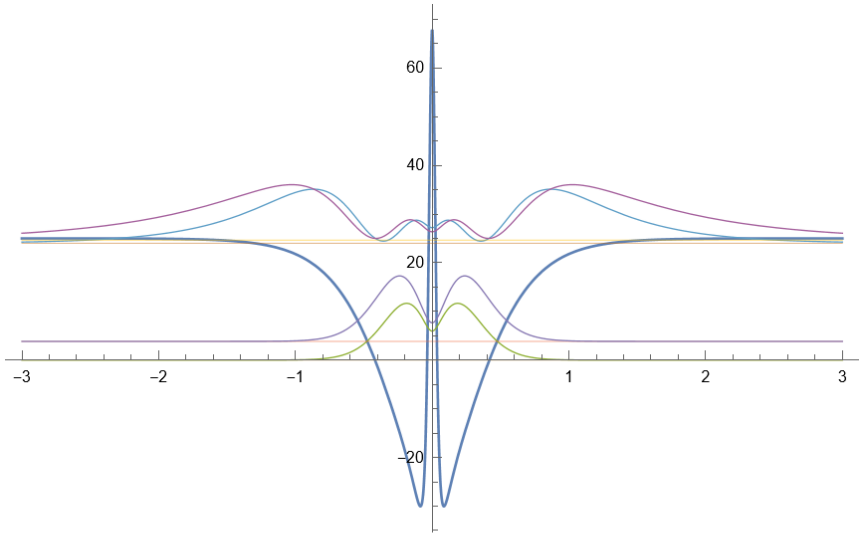

Fig 14. Potential G1301. Only 3 bound states system with tunneling and two band structures, A0=5;B0=6.6;p=4;q=3.35

### 4.13.2 Superpotential $W(x, A_0, B_0) = A_0 \coth(px + iq) - 2B_0 \operatorname{csch}2(px + iq)$

Three cases:
Case 1. The sequence $a_k = (A_k, B_k)$ is defined as follows:

$$A_k = A_0 - 2kp, \tag{173}$$
$$B_k = B_0 - kp \text{ for } k = 0,1,2,.. \tag{174}$$

The eigenvalues are given by
$$E_n = A_0^2 - A_n^2 \tag{175}$$

Under the condition $A_0 > p$, the number of bound states satisfies $1 \leq n < \frac{A_0+p}{2p}$ This potential has triple wells with tunneling between the left and right wells and the central well.

Case 2: The sequence $\{a_k\}$ is defined by
$$A_k = \begin{cases} A_0 - 2B_0 - kp, k \text{ is odd} \\ A_0 - kp, k \text{ is even} \end{cases} \tag{176}$$
$$B_k = B_0 - kp \text{ for } k = 0,1,2,.. \tag{177}$$

The eigenvalues are given by
$$E_n = A_0^2 - A_n^2 \tag{178}$$

The ordering condition on the eigenvalues will require $0 < B_0 \leq \frac{p}{2}$, $A_0 > \frac{1}{2}(2B_0 + 5p)$ with $n = 2k - 1$, the number of bound states is satisfying the condition $1 \leq k < \frac{2A_0 - 2B_0 - p}{4p}$.

This potential has triple wells which can exhibit tunneling through the central forbidden region. See Fig.15.

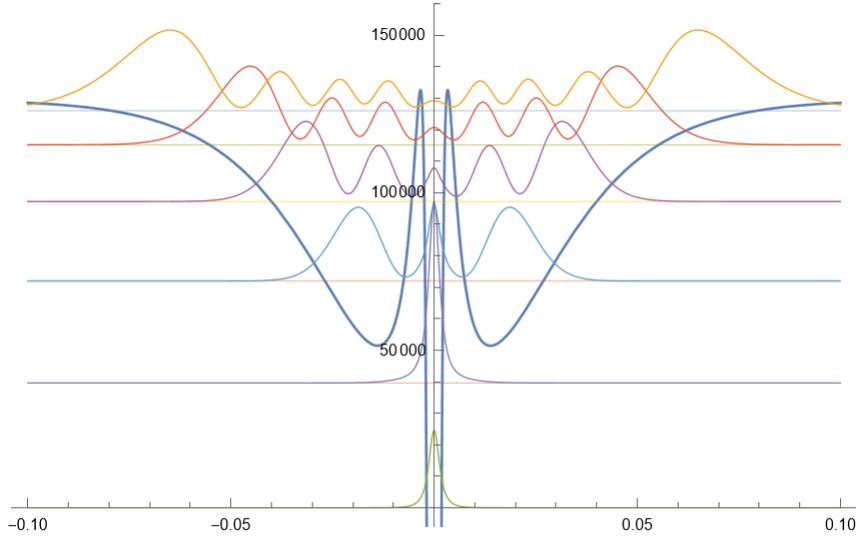

Fig 15. Potential G1302. We observe tunneling between the three wells through the forbidden regions. A0=360,B0=42,p=30,q=4.77

Case 3: The sequence $a_k = (A_k, B_k)$ is defined by
$$A_k = \begin{cases} -A_0 + 2B_0 - kp, k \text{ is odd} \\ A_0 - kp, k \text{ is even} \end{cases} \tag{179}$$
$$B_k = (-1)^k \text{ for } k = 0,1,2,.. \tag{180}$$

The eigenvalues are given by
$$E_n = A_0^2 - A_n^2 \qquad (181)$$

The ordering condition on the eigenvalues will require $B_0 > \frac{5p}{2}$, $\frac{1}{2}(2B_0 - p) < A_0 < \frac{1}{2}(2B_0 + p)$ with $n = 2k - 1$, the number of bound states is satisfying the condition $1 \leq k < \frac{2B_0 - p}{4p}$.

This potential has double wells and triple wells, and tunneling of some states can be observed between them.

# 5  Conclusion

In this paper, the coefficients of the superpotential associated with the Schrödinger equation, as well as the argument, were elevated to the complex plane, and we were able to solve the equation exactly, by exhibiting clearly and explicitly the eigenvalues and the eigenfunctions using the factorization method. Not only does this method help solve the Schrödinger equation, but also the second-order linear differential equation and Riccati equations. It can be seen in some potentials the existence of some properties that were not found in the real case of the known solvable potentials, like the bound states in the continuum. These results also settled the debate about the existence of real eigenvalues when the potential is complex. Furthermore, periodic potentials are easily accessible and multiwells can be constructed from this first proposed list of extended known potentials.

These potentials have prospective applications in different fields of science like physics, chemistry, and biology. Some have a finite number of bound states, while some possess an infinite number, but discrete, of bound states that can be controlled by varying their parameters.

The coming papers will extend the results further to other classified groups, where tunneling and BICs are exhibited.